\documentclass[aps,pra,reprint,showpacs,amsmath,amssymb,groupedaddress,floatfix,nofootinbib] {revtex4-1}

\usepackage{graphicx}

\begin{document}
\title{ Ultracold bosonic scattering dynamics off a repulsive barrier: 
\\ coherence loss  at the dimensional crossover}

\author{V. J. Bolsinger}  \email{vbolsing@physnet.uni-hamburg.de}
\author{S. Kr\"onke}
\author{P. Schmelcher}  \email{pschmelc@physnet.uni-hamburg.de}

\affiliation{Zentrum f\"ur Optische Quantentechnologien, Universit\"at Hamburg, Luruper Chaussee 149, 22761 Hamburg, Germany}
\affiliation{The Hamburg Centre for Ultrafast Imaging, Universit\"at Hamburg, Luruper Chaussee 149, 22761 Hamburg, Germany}
\date{\today}

\begin{abstract}
We explore the impact of dimensionality on the scattering of a small bosonic ensemble in an elongated harmonic trap off a centered repulsive barrier, thereby taking particle correlations into account.
The loss of coherence as well as the oscillation of the center of mass are studied and we analyze the influence of both particle and spatial correlations.
Two different mechanisms of coherence losses in dependence of the aspect ratio are found.
For small aspect ratios, loss of coherence between the region close to the barrier and outer regions occurs, due to spatial correlations, and for large aspect ratios, incoherence between the two density fragments of the left and right side of the barrier arises, due to particle correlations.
Apart form the decay of the center of mass motion induced by the reflection and transmission, further effects due to the particle and spatial correlations are explored.
For tight transversal traps,  the amplitude of the center of mass oscillation experiences a weaker damping, which can be traced back to the population of a second natural orbital, and for a weaker transversal confinement, we detect a strong decay, due to the possibility of transferring energy to transversal excited modes.
These effects are enhanced if the aspect ratio is integer valued.

\end{abstract}

\maketitle

\section{Introduction}

Moving an impurity or obstacle in a superfluid above the critical velocity is well-known to create excitations ~\cite{Landau1940,Feynman1957,Anderson1966,Griffin1996}.
Especially, the dependence on the inter-particle interaction strength and external parameters, such as the flow velocity or trap potentials, have been studied in detail for Bose-Einstein condensates (BEC)\cite{Davis1995}.
For example in one spatial dimension, the bosonic flow past an obstacle \cite{Hakim1997,Leboeuf2001,Pavloff2002}, or sweeping an obstacle  through a BEC have been explored~\cite{Raman1999, Radouani2004, Engels2007, Leszczyszyn2009} including the observation of solitons, chains of solitons and shock waves propagating upstream.
Studies in higher spatial dimensions have been performed as well, yielding the emission of vortices  \cite{Raman1999,Winiecki1999, Winiecki2000}, oblique dark solitons \cite{El2006} and Cherenkov radiation \cite{Carusotto2006}.
Even mesoscopic quantum states have been predicted in these setups for attractive BEC \cite{Streltsov2009, Streltsov2009a, Weiss2009, Gertjerenken2012, Gertjerenken2012a, Helm2012} and a suppression of Cherenkov radiation \cite{Scott2008} if particle correlations are taken into account.

For interferometric setups, inserting 'obstacles'  in the interferometric pathways \cite{Cronin2009} can be used both for splitting and recombining \cite{Berrada2016} the atomic beam or solitons \cite{Martin2012}, similar to light impinging on a half-silvered mirror.
Another possibility of building interferometers is the free-oscillation atom interferometer, where the ground-state wave function in a harmonic trap is excited by a laser pulse into a left and right moving part, which collides again after half an oscillation period similar to a Michelson interferometer  \cite{Wang2005, Garcia2006, Horikoshi2007, Kafle2011, Leonard2012, Fogarty2013}.
When the trapped condensate is initially spatially displaced and impacted by a centered impurity dissipative transport \cite{Dries2010}, dipole oscillations \cite{Albert2008} as well as effects of the inter-particle interactions can been studied \cite{Fogarty2013}.
In both interferometric setups above, a coherent splitting and recombination is important in order to increase the contrast of the interference fringes.
Sources of coherence loss are particle correlations, excitations of transversal modes or coupling to an environment as well as finite temperature.
Usually, the splitting and recombination process of interferometers is describe using a quasi-one dimensionally mean-field approach \cite{Albert2008}, which cannot cover, per construction, loss of coherence, due to particle correlations.
In order to study coherence losses in the complete crossover from three to one dimension, particle correlations and coupling to higher transversal modes have to be taken into account.
Furthermore in interferometers, observables of interest are the probability of reflection  and transmission of the matter wave beam or the oscillation of the center of mass (CM) \cite{Berrada2016}, which show the effectiveness of splitting and recombination.

In the present work, we explore the quantum dynamics of a bosonic ensemble in an elongated trap, which is initially displaced and exposed to by a centered Gaussian barrier, experimentally realizable by a blue-detuned laser beam \cite{Raman1999} or an impurity~\cite{Bonart2013, Catani2012, Schurer2015}.
We vary the aspect ratio of the trap, thereby providing a smooth transition from three to one spatial dimension.
The initial displacement is large enough, such that, by the dimensional coupling of the barrier, higher excited transversal modes can be populated in a controlled manner.
In this way, our work is a natural expansion of our previous studies, where the three dimensional tunneling of few bosons in a double well has been explored at low energies \cite{Bolsinger2017}, or for one-dimensional setups \cite{Schmelcher2008a, Schmelcher2007b}.
We analyze the influence of particle and spatial correlations on the coherence, measured by the first-order correlation function in longitudinal direction, and on the decay of the CM motion.
The strength of the  particle and spatial correlations depend strongly on the aspect ratio of the system.
Two mechanisms of coherence loss are identified.
In the first one, present for low aspect ratios, we observe a loss of coherence between the region close to the barrier and outer regions, due to the excitations of transversal modes.
The second one, at larger aspect ratios, is an incoherence between the density fragments to the right and the left of the barrier,  emerging due to particle correlations and becoming manifest in the occupation of the second natural orbital.
Furthermore, the damping of the CM oscillation is reduced, if particle correlations become dominant, whereas it is enhanced, if spatial correlations are present.
When the aspect ratio is integer valued, we see a quantitative enhancement of these effects for low aspect ratios.

This paper is structured as follows: 
In section \ref{section_Set-Up} the setup and the preparation of the initial state are introduced.
Moreover, we provide the definitions of particle and spatial correlations as employed in this work.
To thoroughly understand the case of  interacting bosons in three dimensions, we proceed in three steps. 
First, presented in section \ref{section_1D_Set-Up}, we focus on few atoms in an one-dimensional trap, where only particle correlations can occur.
Second in section \ref{section_3D_1P}, we study a single atom and change the aspect ratio continuously  to cover the transition from three to one dimension.
We show how the coupling between the dimensions (spatial correlations) reduces the amplitude of the CM oscillation and how the coherence is modified, due to incoherent spherical scattering off the barrier.
If the aspect ratio is integer, the damping of the CM motion and the loss of coherence are enhanced.
In section \ref{section_3D}, we then combine our findings for the few bosons case in the crossover from three to one dimension, taking both particle and spatial correlations into account.
Finally, a conclusion is given in section \ref{section_conclusion}.
A sketch of our numerical method and a discussion of convergence is provided in the appendix \ref{app_MLMCTDHB}.

\section{Set-up and Hamiltonian} \label{section_Set-Up}

We consider the quantum dynamics of $N$ interacting, ultracold bosons each of mass $m$, oscillating in a harmonic trap and scattering off a centered barrier, experimentally realizable by a blue-detuned laser beam or impurity.
The system is governed by the three-dimensional Hamiltonian
\begin{equation} \label{eq_3D_Hamilt}
	\mathcal{H}_{3D}=\sum_{i=1}^N 
	\left[ \mathcal{H}_{0}(\mathbf{r}_i) + \mathcal{V}(\mathbf{r}_i) \right] 
	+ \sum_{1 \leq i < j \leq N} \mathcal{W}(\mathbf{r}_i,\mathbf{r}_j) 
\end{equation}
where $\mathcal{H}_0(\mathbf{r}_i)= -\frac{1}{2} \nabla_{\mathbf{r}_i}^{2} + \frac{1}{2}\left( \eta^{2} x_i^2 + \eta^2 y_i^2 + z_i^2 \right)$ denotes the kinetic energy and trapping potential of the $i$-th atom at position $\mathbf{r}_i= \left( x_i, y_i, z_i \right)^T$ and $\eta= \omega_\bot / \omega_\parallel$ defines the aspect ratio between the transversal $\omega_\bot$ and longitudinal $\omega_\parallel$ trap frequency, which give the characteristic length scales $l_\bot=\sqrt{\hbar / m \omega_\bot}$ and $l_\parallel=\sqrt{\hbar / m \omega_\parallel}$, respectively.
The Hamiltonian is given in dimensionless units, where the energy is scaled w.r.t. $\hbar \omega_\parallel$ and lengths are given in units of $l_\parallel$.
The barrier is described by an external Gaussian potential, $\mathcal{V}(\mathbf{r}_i)= H \exp \left( -r_i^2/S^2 \right)$ 
with height $H$ and width $S=0.2$.
$\mathcal{W}(\mathbf{r}_i,\mathbf{r}_j)$ refers to the three-dimensional short-range interaction between the $i$-th and $j$-th atom.
Since short-range interactions in three dimensions cannot be described properly by a zero-range potential within numerical simulations in the laboratory frame \cite{Bolsinger2017}, a finite-size model potential is needed, which introduces a new length scale to the system, the interaction range $\sigma=0.1$. 
For computational reasons \cite{Bolsinger2017}, we model the interaction by a narrow Gaussian interaction potential $\mathcal{W}_G(\mathbf{r}_i,\mathbf{r}_j) = h \exp \left(-(\mathbf{r}_i-\mathbf{r}_j)^2 / \sigma^{2} \right)$, where $h$ is the height.
The interaction strength can be adjusted by changing $h$, but increasing $h$ leads also to a larger effective interaction range, $\propto \sqrt{h}$, which can penetrate other length scales in the system.
In order to guarantee always short-range interactions, $\sigma < l_\bot \leq 1 $, the Gaussian interaction potential is renormalized w.r.t. a small but arbitrary energy scale of the system $\epsilon$, $\mathcal{W}_G(\mathbf{r}_i,\mathbf{r}_j)(\mathbf{r})=\epsilon$ if $|\mathbf{r}_i-\mathbf{r}_j|=\sigma$, leading to the renormalized Gaussian interaction potential $\mathcal{W}(\mathbf{r}_i,\mathbf{r}_j) = h \exp \left(-(\mathbf{r}_i-\mathbf{r}_j)^2 / \sigma_{eff}^{2} \right)$ with the effective width $\sigma_{eff}=\sigma/\sqrt{\ln(h)}$ and where we set $\epsilon=1$.
In the limit of $h \rightarrow \infty$, we recover the scattering properties of a hard wall interaction model potential with range $\sigma$.

The correspondence between the physically relevant zero-energy scattering length $a_0$ and our interaction model parameters $h$ and $\sigma$ can be obtained by solving numerically the three-dimensional scattering problem in free space and evaluating the limit $E \rightarrow 0$ as investigated in \cite{Bolsinger2017}. 
Due to our renormalization of the Gaussian interaction potential and restricting ourselves to repulsive interaction potentials only, the scattering length is limited for all heights $h$ to $a_0\leq \sigma$.

The initial state is obtained by switching off the barrier $\mathcal{V}$ and letting the system relax to its ground state by an imaginary time propagation via the {\it ab-initio} Multi-Layer Multi-Configuration Time-Dependent Hartree method for Bosons (ML-MCTDHB) (see appendix \ref{app_MLMCTDHB}).
Then the wave function is displaced by a distance $b$ in the elongated direction and instantaneously the barrier is switched on.
The displacement of the wave function is chosen large enough such that its overlap with the barrier is negligible and its gained potential energy is larger than the barrier height $H$, which ensures that we operate in the over-barrier and not in the tunneling regime.
This initial state is then propagated numerically in real time with the ML-MCTDHB method~\cite{Kronke2013, Cao2013, Bolsinger2017, Cao2017}.

We are interested in the influence of the scattering process off the barrier on particle and spatial correlations.
\emph{Particle correlations} are defined as the deviation from a mean-field state, where the wave function can be expressed as a product state w.r.t. the particles, $\Psi(\mathbf{r}_1,...\mathbf{r}_N) = \prod_{i=1}^N \phi(\mathbf{r}_i) $ and \emph{spatial correlations} are given by the deviations from a factorization w.r.t. the spatial directions $\Psi(\mathbf{r}_1,...,\mathbf{r}_N) = \Psi_x(x_1,...,x_N) \Psi_y(y_1,...,y_N) \Psi_z(z_1,...,z_N) $.
In order to quantify the influence of particle and spatial correlations, we use the eigenvalues (natural populations) $a^{(3D)}_i,b_i^{(s)}$ of the reduced density operators  $\rho^{(3D)}=tr_{N-1}|\Psi\rangle \langle \Psi |$ for a single boson in three dimensions and $\rho^{(s)} = tr_{\{ 1,2,3\} \setminus s} \ \rho^{(3D)}$ for the  $s$-th degree-of-freedom of a single boson, respectively.
The first trace stands for an integration over $N-1$ atoms and the second trace is an integration over all but the dimension $s$, respectively.
The natural populations lie in $[0,1]$ and are normalized to one $\sum_i a^{(3D)}_i = \sum_ib^{(s)}_i=1$ and labeled in decreasing sequence.
If only a single eigenvector of $\rho^{(3D)}$ (a so called natural orbital) is populated, $a^{(3D)}_1=1$, all atoms must be in the same orbital, and the many-body wave function can be expressed as a product i.e.  Gross-Pitaevskii mean-field state.
Similarly, deviations of $b^{(s)}_1$ from unity indicate correlations between the spatial dimension $s$ and the other two spatial dimensions.
We note that the employed ML-MCTDHB method can resolve both particle and spatial correlations and can be reduced in a limiting case to a highly efficient solver for the 3D Gross-Pitaevskii Equation (GPE) on large girds (see appendix \ref{app_MLMCTDHB}).

\section{Few bosons ensembles in one dimension} \label{section_1D_Set-Up}

In this section, we analyze the scattering dynamics of a small ensemble of interacting bosons, displaced initially by a distance $b=3$ in a purely one-dimensional harmonic trap off a centered barrier.
First, in section \ref{section_reduction_1D}, we adiabatically reduce the three-dimensional Hamiltonian \eqref{eq_3D_Hamilt} to one-dimension in order to derive the dependence of the corresponding one-dimensional physical parameters on the aspect ratio $\eta$.
Second, in the first part of section \ref{section_1D_results}, we focus on weak interaction strengths as well as moderate barrier heights and describe analytically the collision dynamics by means of a time-dependent two-mode approximation within the mean-field theory.
In the remaining part of the section, the interaction strength and barrier height are increased and we numerically analyze the effect of particle correlations on the oscillation of the CM and the loss of first-order coherence by means of the ab-initio ML-MCTDHB method.

\subsection{Dimensional reduction} \label{section_reduction_1D}

We derive an effective one-dimensional Hamiltonian from the three-dimensional one \eqref{eq_3D_Hamilt}, by integrating out the transversal dimensions, assuming that the energy of the first excited transversal mode is much larger than any other energy scale in our system.
In the limit of large aspect ratios $\eta \rightarrow \infty$, this approach becomes increasingly accurate.
Then the total wave function separates w.r.t.\ the spatial dimensions and the transversal wave function can be described by  all atoms residing in the transversal ground state $\sqrt{\eta/\pi} \exp (-\eta \rho^2)$, where $\rho  = \sqrt{x^2+y^2}$.
This simple reduction is fine for investigating the basic scattering dynamics in a quasi-one-dimensional setting, and we refer the interested reader to the literature~\cite{Kamchatnov2004, Salasnich2002, Tacla2011, Mateo2008, Gerbier2004} for alternative sophisticated methods.
Within our assumptions, the three-dimensional Hamiltonian \eqref{eq_3D_Hamilt} reduces to
\begin{align} \label{eq_1D_Hamilt}
	\mathcal{H}_{1D}=\sum_{i=1}^{N} \left[ H_{0}(z_i) + V(z_i) \right]
	+ \sum_{1\leq i < j \leq N} W(z_i,z_j)
\end{align}
with $H_{0}(z_i)= ( -\partial_{z_i}^2 +  z_i^2)/2$, $V(z_i)=H_{1D} \exp (-z_i^2/S^2)$ and $W(z_i,z_j)=h_{1D} \exp [-(z_j-z_i)^2/ \sigma_{eff}^2]$.
The resulting one-dimensional parameters are $H_{1D}=H\eta S^2/(1+\eta S^2)$ and $h_{1D}=h \eta \sigma_{eff}^2/(2+\eta \sigma_{eff}^2)$.
For a pure one-dimensional setup ($\eta \rightarrow \infty$), the three-dimensional parameters are recovered again, $H_{1D}(\eta \rightarrow \infty)=H$ and $h_{1D}(\eta \rightarrow \infty)=h$.

\subsection{Quantum dynamics in one spatial dimension} \label{section_1D_results}

We analyze the quantum dynamics of five bosons in an one-dimensional trap following the Hamiltonian \eqref{eq_1D_Hamilt}.
In doing so, the interacting ground state is displaced by $b=3$ and gains an additional potential energy of $E=b^2/2=4.5$, which is larger than $H_{1D}$, an thus the bosons reveal dipole oscillations \cite{Brey1989,Kohn1961a,Dobson1994,Fetter1998,Wu2011}, which are modified by the presence of the barrier.
First, we consider weak interactions and small barrier heights, where a mean-field approximation is justified, and then we switch to stronger interactions and larger barrier heights, where particle correlations become important.

\subsubsection*{Small barriers and weak interactions}

\begin{figure}
\centering
\includegraphics[width=\linewidth]{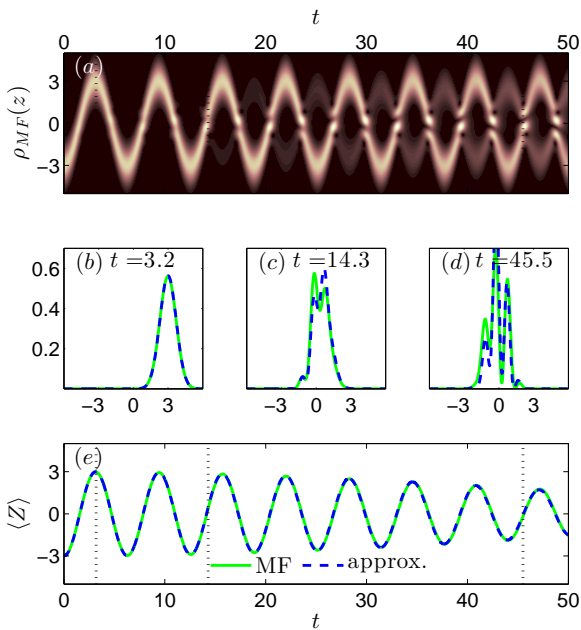}
\caption{ \label{fig_CompareWithModel}
(Color online)
Time evolution of the numerically obtained one-body density (first row) for $N=5$ interacting atoms with $h_{1D}=0.1293$, $\sigma=0.1$ and barrier $H_{1D}=0.4$, $S=0.2$.
The color scales are normalized w.r.t. the maximal value of the density.
(b)-(d) The numerically exact density profile (green line) is compared with the approximation \eqref{eq_ana_n} (blue line) for three instants in time.
In figure (e), the corresponding expectation value of the CM is compared with \eqref{eq_ana_Z}.
Vertical dashed lines mark the three instants in time used in sub figures (b)-(d).
}
\end{figure}

In order to understand the basic scattering behavior, we first focus on five very weakly interacting ($h_{1D}=0.13$, $\sigma=0.1$) bosons and a shallow barrier ($H_{1D}=0.38$, $S=0.2$).
In this regime, a fraction of the initially displaced bosons is reflected at the barrier and cause a counter oscillating wave-packet, which interferes with the transmitted wave-packet (see figure \ref{fig_CompareWithModel}a).
In particular, we do not find any major difference between the results of the {\it ab-initio} ML-MCTDHB simulation and a mean-field calculation, which assumes that all bosons reside in the same orbital $\Psi_{GP}(z,t)$ obeying the one-dimensional GPE
\begin{align} \label{eq_1D_TDGPE}
 & i\partial_{t}\Psi_{GP}(z,t) = \left( H_{0} + V(z) \right) \Psi_{GP}(z,t) + \\
 & + \left( (N-1) \int d\mathcal{Z} |\Psi_{GP}(\mathcal{Z},t)|^2  W(z,\mathcal{Z}) \right) \Psi_{GP}(z,t). \nonumber
\end{align}
where $W(z,\mathcal{Z}) =h_{1D} \exp [-(z-\mathcal{Z})^2/ \sigma_{eff}^2]$.

Furthermore, by inspecting at figure \ref{fig_CompareWithModel}a, we see that the temporal evolution of the density can be modeled by two counter-propagating, stiffly\footnote{\label{footnote1}In the literature, this motion is also called a coherent motion, but in order not to confuse the reader with our other definition of coherence \cite{Glauber1963}, we call it a stiff moving wave packet, since its shape is maintained during time evolution} oscillating wave packets, expressed as two different, stiff modes $\{ \Phi_{i} \}_{i=1,2}$.
For obtaining analytical insights into the dynamics of the density, namely into the interference pattern and the decay of the CM oscillation, we assume that the time-dependent Gross-Pitaevskii orbital $\Psi_{GP}(z,t)$ can be expanded into these two modes 
\begin{equation} \label{eq_1D_expansion}
  \Psi_{GP}(z,t)=A_{1}(t)\Phi_{1}(z,t)+A_{2}(t)\Phi_{2}(z,t)
\end{equation}
As the two modes, we use the solutions of the time-dependent GPE for $V(z)=0$ with the corresponding GPE ground state $\varphi_{GP}$ and energy $ E_{GP}$, being displaced by $\pm b$ as the initial state: $\Phi_{1,2}(z,t) = \exp(-i\Theta(t)) \exp(\pm i\bar{p}(t)z) \varphi_{GP}(z \mp \bar{z}(t)) $ (see appendix \ref{section_1D_TDBF} for the derivation and further details).
Here, $\Theta(t)= E_{GP} t + \frac{1}{2} \bar{z}(t) \bar{p}(t) $ defines the dynamical phase factor and $\bar{z}(t) = b \cos(t) $ as well as $\bar{p}(t) = - b \sin(t)$ are the classical values for position and momentum of an atom oscillating in a harmonic trap.

\begin{figure}[h]
\centering
\includegraphics[width=\linewidth]{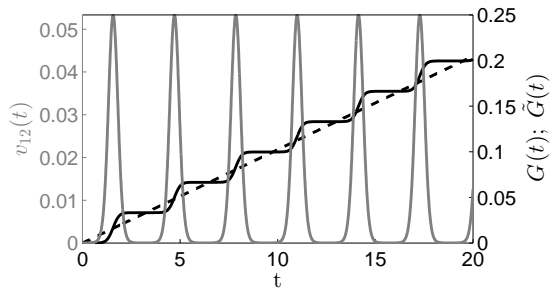}
\caption{ \label{fig_DecayCoef}
Shown is $v_{12}(t)$ (gray solid line), $G(t)$ (black solid line), $\tilde{G}$ (black dashed line) versus time $t$, with $c=0.01$.
$c$ and $v_{12}$  are obtained using the parameters  $\Omega=1$, $H_{1D}=0.38$, $S=0.2$ and $b=3$.
}
\end{figure}

These two modes represent two stiff wave packets, displaced in opposite direction and counter-propagating.
In order to get an analytic expression for $\varphi_{GP}$, we use the Gaussian trial wave function $(\Omega/\pi)^{1/4} \exp (-\Omega z^2/2 )$, and determine the parameter $\Omega$, which incorporates the effect of the interaction, by minimizing the total energy.
For a non-interacting system, $\Omega$ is equal to one, and increasing the repulsive (attractive) interaction leads to a broadening (narrowing) of the trial wave function and thus to a smaller (larger) $\Omega$. 
Knowing the functional expression for $\varphi_{GP}$, the overlap between $\Phi_{1}(z,t)$ and $\Phi_{2}(z,t)$ can be calculated analytically and  equals $\exp(-b^2 (\Omega^2 \cos^2(t)+\sin^2(t))/ \Omega )$. 
For large displacements $b$, as regarded here, these two modes are approximately orthogonal for all times, of which we will make use in the following.
Inserting the expansion \eqref{eq_1D_expansion} into the time-dependent GPE \eqref{eq_1D_TDGPE}, projection onto the two modes, $\Phi_{i=1,2}$, and employing the 
symmetries  $\varphi_{GP}(z)=\varphi_{GP}(-z)$ 
and $V(z)=V(-z)$, leads to the equations of motion for the coefficients 
\begin{equation} \label{eq_1D_coupled_Modes}
 i \partial_t  \left( \begin{array}{c} A_1 \\ A_2 \end{array} \right) 
 =\left( \begin{array}{cc} v_{11} & v_{12}\\ v_{12} & v_{11} \end{array} \right) 
 \left( \begin{array}{c} A_1 \\ A_2 \end{array} \right)
\end{equation}
where the matrix elements are ${v_{11}={\int dz |\varphi_{GP}(z-\bar{z}(t))|^2 V(z)}}$ and $v_{12}= {\int dz \exp (-i2 \bar{p}(t)z)}   {\varphi_{GP}(z-\bar{z}(t))}  {\varphi_{GP}(z+\bar{z}(t)) V(z) }$.
These last two integrals can be evaluated, and yield
\begin{align}
v_{11}(t)  &=
\frac{\sqrt{\Omega S^2}}{\sqrt{1+\Omega S^2}}    H_{1D}
e^{ - \frac{\Omega b^2 \cos^2(t)}{1+\Omega S^2}} \\ 
v_{12}(t) & = 
\frac{\sqrt{\Omega S^2}}{\sqrt{1+\Omega S^2}}   H_{1D}
e^{ - b^2 \left( \Omega  \cos^2(t) + \frac{S^2 \sin^2(t)}{1+\Omega S^2}   \right)} \nonumber
\end{align}
The set of equations \eqref{eq_1D_coupled_Modes} can be solved analytically 
\begin{align} \label{eq_1D_solution}
  A_{1}(t) &= e^{-iF(t)}\cos(G(t)) \\
  A_{2}(t) &= -ie^{-iF(t)}\sin(G(t)) \nonumber
\end{align}
with $F(t)=\int_{0}^{t}v_{11}(\tau)d\tau$ and $G(t)=\int_{0}^{t}v_{12}(\tau)d\tau$. 
Since the modes $\Phi_{1,2}$ couple only during the collisions, $v_{12}(\tau)$ is periodic and strongly peaked such that $G(t)$ increases step-like  (see figure \ref{fig_DecayCoef}).
In order to simplify $G(t)$, we apply a stationary phase approximation for $v_{12}$ piecewise in the time intervals $[n\pi,(n+1)\pi)$, with $n\in \mathbb{N}_0$, and furthermore perform a linear fit $G(t)\approx\tilde{G}(t)=ct$ with
\begin{equation} \label{eq_ana_c}
c = \frac{1}{\sqrt{\pi}b} \sqrt{\frac{\Omega s }{\Omega-s}} H_{1D} e^{-b^2 s} 
\end{equation}
where $s=S^2/(1+\Omega S^2)$.
These approximations are quite accurate as can be seen in figure \ref{fig_DecayCoef}.
Calculating the evolution of the  CM $\langle Z \rangle = \sum_{i=1}^N \langle z_i  \rangle /N$
\begin{align}
 \langle Z \rangle = & b \cos(t) \cos(2ct)                           \label{eq_ana_Z}
\end{align}
we find that the classical oscillation of a displaced atom in a harmonic trap ($\langle Z \rangle = b \cos(t)$) is modified by a slower oscillation $\propto \cos(2ct)$, causing a decay and revival of $\langle Z \rangle$.
Thus $c$ determines the time-scale, on which the classical CM oscillation 'decays', namely $t_d=\pi/(4c)$, and we therefore called it \emph{decay coefficient} in the following.
The decay coefficient $c$ depends strongly on $b$, and if $b$ is increased, $c$ reduces towards zero, meaning that if the initial wave function is stronger displaced, it has more kinetic energy, travels faster through the barrier and thus the coupling with the barrier is reduced.
Similarly, decreasing the barrier height $H_{1D}$, the coupling to the barrier is reduced, and $c \rightarrow 0$, leading to an undamped CM oscillation, $\langle Z \rangle = b \cos(t)$.
The influence of the interaction strength on $c$ is rather small in the weak interacting regime, which we address with the Gaussian trial function for $\varphi_{GP}$,  and in the validity of our model, $c$ can be assumed as constant,  $c(h_{1D})\sim const$.
Furthermore, the density $|\Psi(z,t)|^2$ can be calculated analytically
\begin{align}
  & |\Psi(z,t)|^2 =  \cos^2(ct) |\varphi_{GP}(z-\bar{z})|^2   \nonumber \\
  & + \sin^2(ct) |\varphi_{GP}(z+\bar{z})|^2       \label{eq_ana_n} \\
      &  + \sin(2 \bar{p} z) \sin(2ct) \varphi_{GP}(z-\bar{z}) \varphi_{GP}(z+\bar{z})    \nonumber
\end{align}
where we have omitted the time-argument for $\bar{z}$ and $\bar{p}$ for better readability.
The density \eqref{eq_ana_n} consists of three parts. The first two terms describe stiff, out-of-phase oscillations of the ground state wave functions $|\varphi_{GP}(z-\bar{z})|$ and $|\varphi_{GP}(z+\bar{z})|$ with a sinusoidal population transfer of frequency $2c$ between these two states.
The last term creates an interference pattern with a contrast $\propto\sin(2ct)$ and is strongest at $t=t_d+n\pi/2c$, $n\in\mathbb{N}_0$.
In the limit of $c\rightarrow 0$, this interference pattern blurs and one is left with a stiff oscillation of a single wave packet.

For different instants in time, we compare the approximate solution \eqref{eq_ana_n} for the density and \eqref{eq_ana_Z} for the CM with the full mean-field calculation (figure \ref{fig_CompareWithModel}).
Not only the decay of the CM $\langle Z \rangle$ oscillations is well-described by our simple model, but also the interference pattern. 
Nevertheless, let us finally discuss the implicit assumptions underlying our analytical approach.
Deviations in our model occur, since the barrier can scatter atoms into higher excited modes, which are not taken into account by the model.
Therefore, the model is only valid for small to moderate barrier heights.
Furthermore, while our model assumes elastic scattering off the barrier, the scattering is inelastic in fact, which can be seen in the mean-field calculations showing that the turning points of the reflected density fraction are closer to the trap center than for the transmitted density fraction (see figure \ref{fig_CompareWithModel}a).
Finally, the assumption of a Gaussian trial wave function limits the model to small interactions and, summarizing, we find empirically that the 'decay' of the CM is slightly decreased in fact if the interaction strength is increased, whereas our model features the opposite trend.

\subsubsection*{Large barrier amplitude and stronger interactions}

For stronger interactions $(h_{1D}=1.5385)$ and larger barrier height $(H_{1D}=1.5)$, the dynamics cannot be described by the GPE \eqref{eq_1D_TDGPE} anymore, since correlations between the atoms have to be taken into account.
Quantitative differences between a MF and a BMF simulation are observed, for example, in the oscillation of the CM, in the interference pattern of the density or in the one-body correlation function.
In the following, we explain the occurrence of these quantitative differences, starting with the interference pattern.

\begin{figure}[h]
\centering
\includegraphics[width=\linewidth]{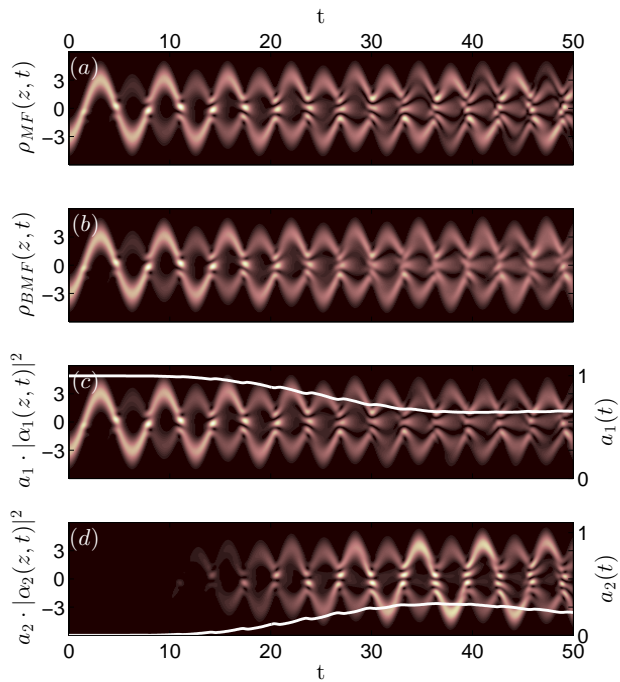}
\caption{ \label{fig_1D_BMF_Density}
(Color online)
Temporal evolution of the density both within  the 1D-GPE  (a) and when taking particle correlations into account (b), for  $N=5$ bosons in a harmonic trap with barrier $H_{1D}=1.54$, $S=0.2$ and interaction $h_{1D}=2.01$, $\sigma=0.1$.
(c) and (d) show the first and second natural orbitals $\alpha_{1,2}(z,t)$ weighted with their natural populations $a_{1,2}$, which are shown as white solid lines. 
The color scales are normalized w.r.t. the maximal value of the density.
}
\end{figure}

While the MF density $\rho_{MF}(z,t)$ reveals a pronounced interference pattern for all times (see figure \ref{fig_1D_BMF_Density}a), the interference pattern becomes smeared out as time proceeds if particle correlations are taken into account (see figure \ref{fig_1D_BMF_Density}b).
The reason for the loss of coherence is a significant depletion of the first natural orbital $\alpha_1(z,t)$ down to 65\% of the original population (white line in figure \ref{fig_1D_BMF_Density}c), which mainly stems from populating the second natural orbital $\alpha_2(z,t)$  up to 29\% (white line in figure \ref{fig_1D_BMF_Density}d).
The remaining 6\% are distributed among the remaining four higher natural populations (not shown).

\begin{figure}[h]
\centering
\includegraphics[width=\linewidth]{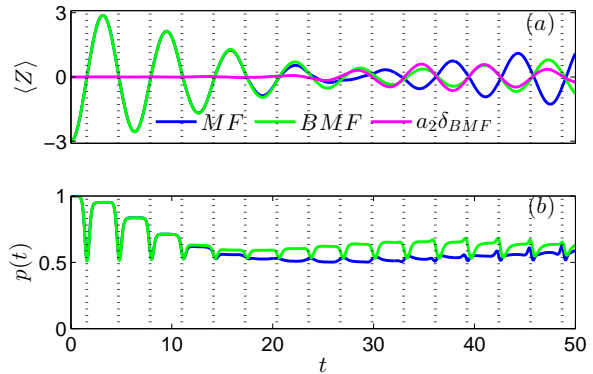}
\caption{ \label{fig_1D_BMF_COM}
(Color online)
Sub figure (a) shows the CM oscillation for and sub figure (b) shows the probability to find two bosons on the same side of the trap, see equation \eqref{eq_1D_p}.
Same parameters as in figure \ref{fig_1D_BMF_Density} are used.
For both figures, the dark blue line represents the mean-field result \eqref{eq_1D_TDGPE} and the green light line denotes a ML-MCTDHB result, where particle correlations are taken into account.
The magenta line represents the expectation value of $\delta_{BMF}=\langle z \rangle$ evaluated w.r.t. the second natural orbital and weighted with its  population.
}
\end{figure}

We can approximate the BMF density $\rho_{BMF}$ by the two main contributing natural orbitals, $\rho_{BMF} \approx a_1|\alpha_1(z,t)|^2 + a_2|\alpha_2(z,t)|^2$, with $a_1+a_2 \approx 1$.
Empirically, the first natural orbital has a similar structure as the mean-field density (compare sub figures \ref{fig_1D_BMF_Density}a and \ref{fig_1D_BMF_Density}c), $|\alpha_1(z,t)|^2 \sim \rho_{MF}$, and the BMF expectation value for the CM can be approximated as
\begin{equation} \label{eq_Z_approx}
\langle Z \rangle_{BMF} \sim (1-a_2) \langle Z \rangle_{MF} + a_2 \int dZ \; Z |\alpha_2(Z,t)|^2
\end{equation}
Inspecting figure \ref{fig_1D_BMF_COM}a, where we show the second part $ \delta_{BMF}=\int dZ \; Z |\alpha_2(Z,t)|^2$ multiplied by $a_2$, we notice that $\delta_{BMF}$ is in-phase with the classical harmonic oscillation.
By increasing $a_2$ this in-phase oscillation becomes more pronounced in $\langle Z \rangle_{BMF}$ and therefore the decay of the CoM in reduced and $t_d$ is increased.
This effect can be also seen in figure \ref{fig_1D_BMF_COM}, where both $\langle Z \rangle_{BMF}$ and $\langle Z \rangle_{MF}$ are presented.

Next, we study the probability for two bosons being on the same side of the barrier, which effectively measures the probability of pairwise transmission/reflection at the barrier
\begin{align} \label{eq_1D_p}
p(t)=& \frac{1}{N(N-1)} \nonumber \\
& \sum_{1\le i < j \le N } \left( \langle \Theta(z_i) \Theta(z_j) \rangle + \langle \Theta(-z_i) \Theta(-z_j) \rangle \right)
\end{align}
where $\Theta$ is the Heaviside function.
The probability $p(t)$ is enhanced, if particle correlations are taken into account (see figure \ref{fig_1D_BMF_COM}b), identifying pair correlation.
So, the bosons like to be transmitted or reflected pairwise.
Pair correlations have already been observed in a double well scenario \cite{Schmelcher2008c}, similar to our setup, but focusing on tunneling dynamics.
This feature of enhanced pair correlation can also be seen in the two-body density matrix (not shown) and, as we shall see next, causes a decrease of the one-body coherence in the dynamics, which is consistent with the disappearance of the interference pattern in the temporal evolution of the density.

\begin{figure}[h]
\centering
\includegraphics[width=\linewidth]{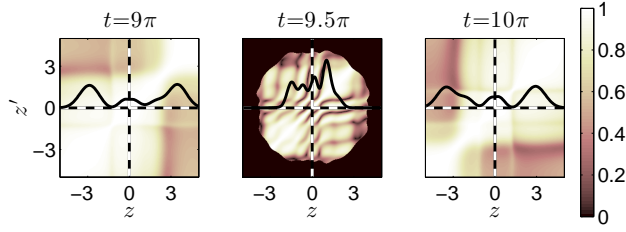}
\caption{ \label{fig_1D_g1}
Absolute value of the first-order correlation function for three different times $t=\{9\pi, \; 9.5\pi, \; 10\pi \}$.
The blacks line indicate the density profiles at the corresponding time instants and the black-white dashed lines mark the position of the barrier.
Areas of density smaller than $10^{-6}$ are colored in black.
}
\end{figure}

Finally, we analyze how the emergent particle correlations affect the spatial coherence of the bosonic ensemble by inspecting the first-order correlation function 
\begin{equation} \label{eq_g}
g_1(z,z')=\rho_{1D}(z,z') / \sqrt{ \rho(z) \rho(z')}
\end{equation}
where $\rho_{1D}$ is the one-dimensional one-body density matrix and $\rho(z)=\rho_{1D}(z,z)$ the one-dimensional one-body density.
The first-order correlation function equals unity in a MF simulation and features values $|g_1(z,z')| < 1$ if particle correlations are present. 
In figure \ref{fig_1D_g1}, $|g_1(z,z')|$ is given for three different times $t= 9\pi$, $9.5\pi$ and $10\pi$.
The first and the last instant in time correspond to the $9$-th and $10$-th classical turning point of the CM oscillation and $t=9.5\pi$ refers to the tenth collision with the barrier.
At the classical turning points, we find that the coherence between the density fragments to the right and the left of the barrier has been reduced due to the depletion of the dominant natural orbital such that the interference contrast is reduced at the subsequent collision. At the collision times, however, the coherence function features an involved ripple structure, which is difficult to interpret. 

\section{One atom in three dimensions} \label{section_3D_1P}

In this section, we analyze the scattering dynamics of a single  atom at the barrier in the crossover from three to one spatial dimension by varying the aspect ratio in order to focus solely on the impact of spatial correlations without any particle correlations.
The atom is initially displaced by $b=3$ in the longitudinal dimension ($z$-direction i.e. $s=3$) and oscillates longitudinally back and forth and scatters (approximately spherically) at the centered barrier $(H=5)$.
First, we inspect the temporal evolution of the density in both transversal and longitudinal direction, and then discuss the effect of  integer and non-integer aspect ratios on the participating scattering channels, followed by an analysis of the CM oscillation and the loss of longitudinal first-order coherence in the system.

\begin{figure}[h]
\centering
\includegraphics[width=\linewidth]{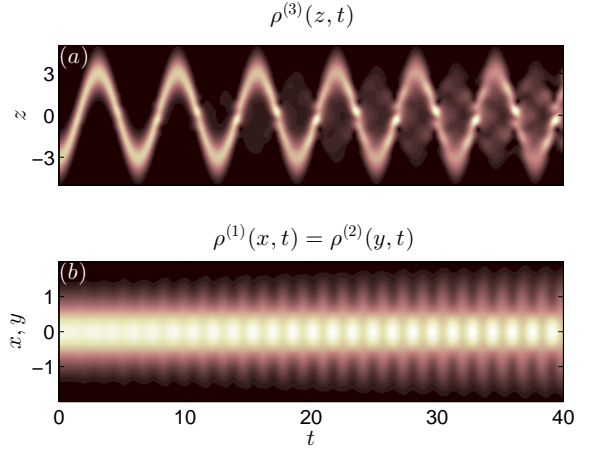}
\caption{ \label{fig_3D_1P_Dens}
(Color online)
Figures (a) and (b) show the longitudinal and transversal density profiles, respectively.
Parameters are:  barrier height $H=5$, width $S=0.2$, displacement $b=3$ and aspect ratio $\eta=2$.
The color scales are normalized w.r.t. the respective maximal value of the density.
}
\end{figure}

In figures \ref{fig_3D_1P_Dens}a and \ref{fig_3D_1P_Dens}b, we depict the time-evolution of the longitudinal and transversal density profiles, respectively.
No interference pattern is observed in the longitudinal density profile and the density becomes much more delocalized as well as more irregular compared to the one-dimensional simulations (cf. figure \ref{fig_1D_BMF_Density}).
While in one dimension, the atom can only be transmitted or reflected, in three dimensions also transversal modes may be populated, since the barrier induces a coupling between the longitudinal and transversal modes. 
The transversal excitations manifest themselves as a breathing of the density (see figure \ref{fig_3D_1P_Dens}b) \cite{Schmitz2013, Chevy2002, Abraham2014}.
Scanning the aspect ratio, we empirically find that the transversal breathing excitations are enhanced if the aspect ratio is integer valued, i.e.\ $\eta\in\mathbb{N}$.
The mechanism of breathing mode excitation is discussed later in detail.

In order to analyze the channels participating in the scattering process as well as the influence of the aspect ratio, we project the numerically ML-MCTDHB obtained wave function $\Psi(\mathbf{r},t)$ onto the following co-moving basis
\begin{equation} 
  \phi_{n,l,m}(\mathbf{r},t) = \varphi^{2D}_{n,l}(\rho, \theta) \phi_m(z,t)
  \nonumber
\end{equation}
which are products of the one-dimensional, periodically moving, stiff wave functions  $\phi_m(z,t)$ multiplied by static transversal harmonic oscillator eigenfunctions $ \varphi^{2D}_{n,l}(\rho, \theta)$.
For the complete orthonormal basis states in the longitudinal direction, we take the solutions $\phi_m(z,t)$ of the time-dependent Schr\"odinger equation for an one-dimensional harmonic oscillator with the $m$-th harmonic oscillator eigenstate  $\varphi^{1D}_{m}$  initially at rest and displaced by $b$ as the corresponding initial condition  (for a derivation see appendix \ref{section_1D_TDBF}).
The stiff wave functions  $\phi_m(z,t)$ have the following functional form $\phi_m(z,t)=e^{-i \Theta_m(t)} e^{+i \bar{p}(t)z} \varphi^{1D}_{m} (z-\bar{z}(t))$, with $\Theta_m(t)=E_m t + \frac{1}{2} \bar{z}(t) \bar{p}(t)$ and the harmonic oscillator eigenenergies $E_m$.
Both $\bar{z}(t)=b \cos(t)$ and $\bar{p}(t)=- b \sin(t) $  are the classical values for the position and momentum of an atom oscillating in a harmonic trap.

\begin{figure}[h]
\centering
\includegraphics[width=\linewidth]{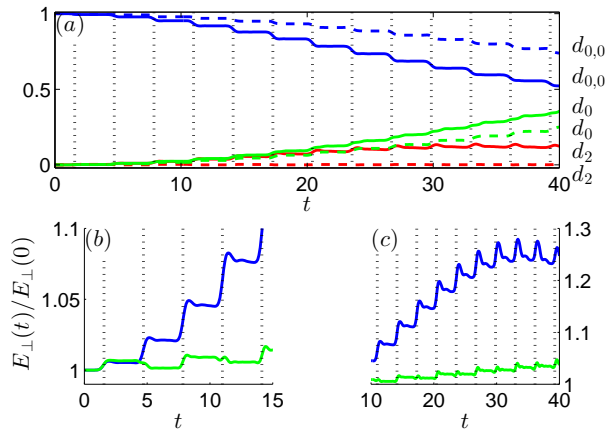}
\caption{ \label{fig_3D_Coef}
(Color online)
(a) Populations of $d_{0,0}$ (dark blue), $d_{2}=\sum_{m=0}^\infty d_{2,m}$ (red), $d_{0}=\sum_{m=1}^\infty d_{0,m}$ (light green) for the aspect ratio $\eta=3$ (solid lines) and for $\eta=2.5$ (dashed lines).
(b) Normalized transversal energy $E_\bot(t)/E_\bot(0)$ for the integer aspect ratio $\eta=3$ (dark blue solid line) and the half-integer aspect ratio (light green solid line) for the first oscillation periods, continued in sub figure (c) for longer times.
Black dotted lines mark times, when the atom scatters with the barrier.
All other parameters as in figure \ref{fig_3D_1P_Dens}.
}
\end{figure}

The initial state of the problem at hand is the Gaussian ground state displaced  by $\bar{z}(0)=b$ in the longitudinal direction, i.e.  $\Psi(\mathbf{r},0)=\phi_{0,0,0}(\mathbf{r},0)$.
Without the barrier, the stiff Gaussian wave packet $\phi_{0,0,0}(\mathbf{r},t)$, oscillating in the longitudinal direction, would  exactly coincide with the solution of the time-dependent Schr\"odinger equation.
In contrast to this, the barrier $\mathcal{V}$ couples various $\phi_{n,l,m}(\mathbf{r},t)$ while respecting the following symmetry-induced selection rule.
The Hamiltonian $\mathcal{H}_{3D}=\mathcal{H}_0+\mathcal{V}$  commutes with  the $z$-component of the angular momentum operator $L_z$. 
Since $\phi_{n,l,m}(\mathbf{r},t)$ is an eigenstate of $L_z$ with eigenvalue $l$, which holds, in particular, for the initial state $\phi_{0,0,0}(\mathbf{r},0)$ with $l=0$, the barrier may only couple states with vanishing angular quantum number, i.e.\ $\phi_{n,0,m}(\mathbf{r},t)$.
In order to monitor both transversal excitations and deviations from the stiff Gaussian wave packet oscillation, we show the probabilities
\begin{equation}
  d_{n,m}(t) \equiv | \langle\Psi(t)|\phi_{n,0,m}(t)\rangle |^2
\nonumber
\end{equation}
for an integer aspect ratio $\eta=3$ (solid lines) and a non-integer aspect ratio $\eta=2.5$ (dashed lines) in figure \ref{fig_3D_Coef}a.
If the aspect ratio is integer valued, the population of the mode $\phi_{0,0,0}$ is transferred both to the second excited transversal harmonic oscillator state $\varphi_{2,0}^{2D}$ measured by $d_2\equiv\sum_{m=0}^\infty d_{2,m}$  as well as to higher excited longitudinal states in the comoving frame with the transversal degrees of freedom being in the ground state, measured by  $d_0 \equiv \sum_{m=1}^ \infty d_{0,m}$, which destroys the stiff oscillation of the wave function.
We find that (for our parameter values) essentially no other states participate in the dynamics, i.e.\ $d_{0,0}+d_{2}+d_{0}\approx1$, because the first excited transversal state cannot be excited for symmetry reasons and  the excitation energy is not sufficient to populate even higher transversal modes.
In the considered time interval, the population $d_{2}$ saturates, whereas the population of $d_{0}$ monotonously increases.
Since these higher excited longitudinal modes are more delocalized, the density becomes more delocalized, too (see figure \ref{fig_3D_1P_Dens}).

In contrast to the integer valued case, no significant population of  $\varphi_{2,0}^{2D}$ can be observed  for the non-integer aspect ratio $\eta=2.5$, in favor of $d_{0}$.
In total the loss of population of the mode $\phi_{0,0,0}$ is weaker for non-integer aspect ratios.
Thus the shape of the wave function is 'stiffer' and the density is less delocalized (not shown) for non-integer aspect ratios.

The suppression of transversal excitations for non-integer valued $\eta$ can be explained by a simple multiple-scattering model,  focusing on the transversal degrees of freedom only and assuming instantaneous collisions with the barrier: 
At the first collision ($t_1=\pi/2$) of the wave function with the barrier, the second transversal mode $\varphi_{2,0}^{(2D)}$ is excited, since the first transversal mode cannot be populated due to symmetry constraints.
Directly afterwards, the time evolution is governed by $\mathcal{H}_0$ only.
Neglecting correlations between the spatial directions induced by the scattering, the time-dependent transversal wave function for $t\in (\pi/2, 3\pi/2]$ is therefore given by
\begin{equation}
   b_{00} \varphi_{0,0}^{(2D)} + b_{02} e^{-i2\eta (t-t_1)} \varphi_{2,0}^{(2D)}  \nonumber
\end{equation}
and up to a global phase factor with the amplitudes $b_{00}$ and $b_{02}$, which determine the transitions $\varphi_{0,0}^{(2D)} \rightarrow \varphi_{0,0}^{(2D)}$ and $\varphi_{0,0}^{(2D)} \rightarrow \varphi_{2,0}^{(2D)}$, respectively.
This excitation leads to a transversal breathing in the density with frequency $2\eta$.
After the time interval $\Delta t = \pi$, the wave function collides with the barrier again, leading to a new excitation from the ground state to the second transversal mode and vice versa.
Excitations to higher modes are neglected again.
At this instant in time ($t_2=3\pi/2$), two additional scattering processes have to be taken into account $\varphi_{2,2}^{(2D)}\rightarrow\varphi_{2,0}^{(2D)}$ and $\varphi_{2,2}^{(2D)}\rightarrow\varphi_{2,2}^{(2D)}$ with the amplitudes $b_{20}$ and $b_{22}$, respectively.
The time-dependent transversal wave function is then
\begin{align}
 B_0 \varphi_{0,0}^{(2D)} +  B_2 \varphi_{2,0}^{(2D)} \nonumber
\end{align}
with $B_0=b_{00}b_{00} + b_{02}b_{20}e^{-i 2 \eta \pi}$ and $B_2= b_{00}b_{02}(1+e^{-i2\eta \pi} b_{22}/b_{00})$.
Within first-order time-dependent perturbation theory, the amplitudes $b_{22} $ and $ b_{00}$ are of the same order $b_{22} \simeq b_{00}$ and have the same phase relation.
We approximate $B_2 \simeq 2 b_{02} b_{00} $ for an  integer aspect ratio $\eta=n$ and $B_2 \simeq 0$ for  the half integer aspect ratio $\eta=(2n+1)/2$, with $n\in \mathbb{N}_0$. 
In other words, the breathing excitations induced by the first and the second collision interfere constructively for integer aspect ratios, whereas they interfere destructively for half-integer valued $\eta$.

This feature can also be clearly seen in the transversal energy,  $E_\bot=\langle -(\partial_x^2 +\partial_y^2 )/2+ \eta^2(x^2+y^2)/2 \rangle$ (see figure \ref{fig_3D_Coef}b,c).
At the first scattering event off the barrier ($t_1=\pi/2$), longitudinal kinetic energy is transformed into transversal excitation energy both for the integer and non-integer aspect ratio.
But at the second scattering event at $t_2=3\pi/2$, the transversal excitation energy is reduced again for the non-integer valued case, whereas in the integer valued case more energy is deposited transversally.
This effect causes the step like structure in $E_\bot(t)$ for integer aspect ratios for times $t<20$.
As times goes by, the wave function becomes more and more delocalized and the matrix element $\langle \Psi(t) | V^{(1)} | \Psi (t) \rangle$ couples the transversal and longitudinal dimensions not only at the main scattering events ( i.e.\ at $t=(2n+1)/2 \; \pi$ with $n \in \mathbb{N}_0$) but all the time.
Thus, following a main scattering event, where energy is pumped into the transversal degrees of freedom, energy can 'flow' continuously back to the longitudinal degree of freedom.
This causes the change from the step-like transversal energy increase to a peak-like one (see figure \ref{fig_3D_Coef}c).
Summarizing, varying the number of oscillation events with the barrier and the aspect ratio might be used for preparing the atom in a certain state involving longitudinal and transversal excited modes.

\begin{figure}[h]
\centering
\includegraphics[width=\linewidth]{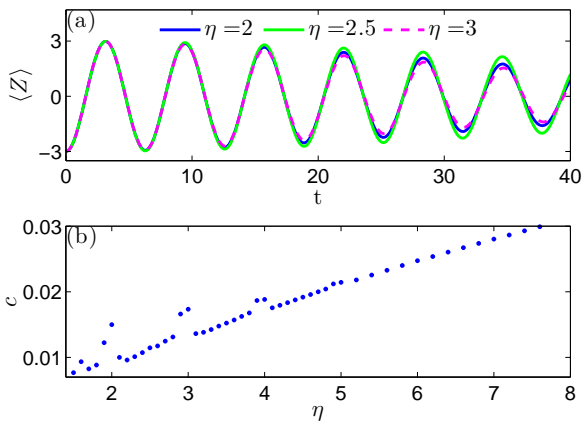}
\caption{ \label{fig_1P_3D_COM}
(Color online)  (a) The oscillation of the CM for $\eta=2$ (dark blue line), $\eta=2.5$ (light green line) and $\eta=3$ (magenta dashed line).
(b)  depicts the fitted decay coefficient  $c$  of model \eqref{eq_ana_Z} for various aspect ratios. 
All other parameters coincide with those of figure \ref{fig_3D_1P_Dens}.
}
\end{figure}

Due to the spatial coupling (in the case of a low integer aspect ratio), we expect also a modification of the CM oscillation, since the barrier can transfer longitudinal kinetic energy into transversal energy, inducing this way a decay mechanism for the longitudinal CM oscillation $\langle Z \rangle$.
In figure \ref{fig_1P_3D_COM}a, we show the CM oscillation for $\eta\in\{2, 2.5, 3\}$ and observe that the CM oscillations decay faster for the integer cases, where significant excitations of  the transversal mode are possible.
In order to analyze the influence of the aspect ratio on the decay of the CM oscillations, we fit the model \eqref{eq_ana_Z} to our numerical data and extract the decay coefficient $c$ (see figure \ref{fig_1P_3D_COM}b).
If the aspect ratio is integer valued, $c$ is peaked, indicating the mentioned decay mechanism w.r.t. the transversal excitation.
These peak heights are reduced for larger aspect ratios since a higher initial excitation energy would be needed to populate the transversal modes.
For even larger aspect ratios, $c$ saturates and corresponds to a pure one-dimensional simulations with the effective physical parameters stated in section \ref{section_reduction_1D} (not shown).

\begin{figure}[h]
\centering
\includegraphics[width=\linewidth]{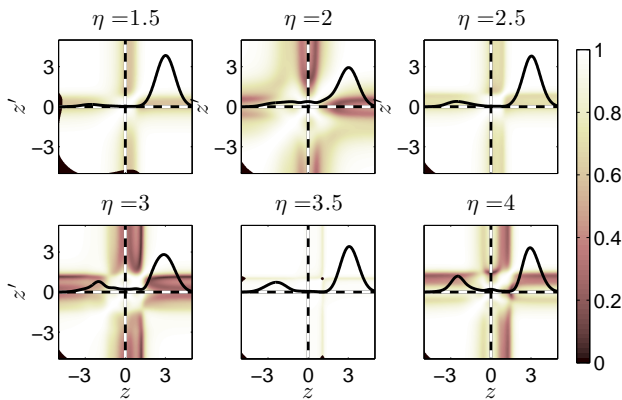}
\caption{ \label{fig_1P3D_g1}
First-order correlation function $|g_1(z,z')|$ for different aspect ratios $\eta$ at the time instant $t=9\pi$.
Areas of density smaller than $10^{-6}$ are colored in black.
The black-white dashed line marks the position of the barrier and the black solid line is the longitudinal density profile.
Same physical parameters are used as in figure \ref{fig_3D_1P_Dens}.
}
\end{figure}
To analyze the loss of coherence in the longitudinal direction due to correlations between the spatial directions, we compare the first-order correlation function 
\begin{equation}
g_1(z,z')=\frac{\rho^{(s=3)}(z,z')}{\sqrt{\rho^{(s=3)}(z)\rho^{(s=3)}(z')}} \nonumber
\end{equation}
where the longitudinal one-dimensional density $\rho^{(s=3)}(z')$ and one-dimensional density matrix $\rho^{(s=3)}(z,z')$ are obtained by integrating out the transversal degrees of freedom, e.g.\ $\rho^{(s=3)}(z,z')=\int dx dy \; \rho^{(3D)}(x,y,z;x,y,z')$.
The absolute value of the first-order correlation function is shown in figure \ref{fig_1P3D_g1} for different aspect ratios at the turning point $t=9 \pi$ of the corresponding classical oscillation.
For integer aspect ratios, we find a pronounced loss of coherence between the region close to the barrier and outer regions.
This incoherent density fraction stems from nearly spherical, incoherent scattering (involving the second excited transversal mode) off the barrier. 
Due to a stronger coupling of the spatial directions, this loss of coherence is enhanced for decreasing $\eta$.
Accordingly, there is only a faint incoherent density fraction for non-integer aspect ratios, being hardly visible in the case $\eta=3.5$.
Non-integer aspect ratios are thus favorable if one needs to propagate an initial wave function coherently w.r.t.\ the longitudinal direction in the presence of a perturber or impurity.
We finally remark that  $g_1(z,z')$ contains for large aspect ratios an asymmetry w.r.t. the barrier in comparison with smaller aspect ratios at the time instant $t=9 \pi$. 
A symmetric arrangement of  $g_1$ around the barrier can be found at an earlier time instant in which the time shift increases.

\section{Few bosons in three dimensions} \label{section_3D}

We combine now the knowledge, which we have gained for a few atoms in one dimension, with one atom in three dimensions, in order to study few atoms in three dimensions taking into account correlations.
We simulate the temporal evolution of five interacting bosons, ($h=91.125$) with a centered barrier ($H=9$) and vary the aspect ratio $\eta$ between $1.5$ and $8.0$, in order to see the influence of the dimensionality on the scattering behavior.
In order to ensure short range interactions and to resolve the interaction potential properly, a large number of grid points have to be used.
The ML-MCTDHB method (see appendix \ref{app_MLMCTDHB}) is tailored to effectively treat large number of grid points and we perform our simulations with 800 and 200 grid points in the longitudinal and each transversal direction, respectively.
For a convergence study and other numerical parameters, see also appendix \ref{app_MLMCTDHB}.

\begin{figure}[h] 
\centering 
\includegraphics[width=\linewidth]{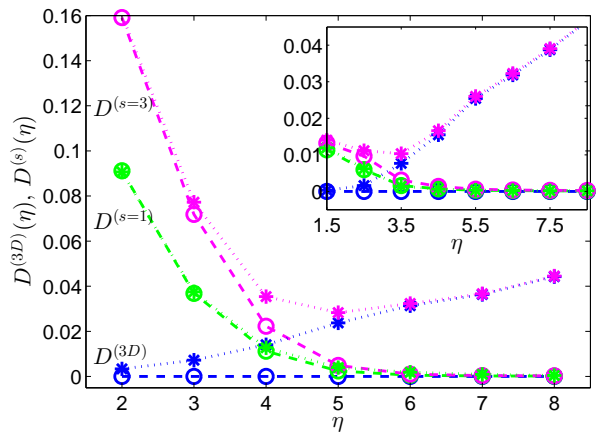}
\caption{ 
(Color online)
Time-averaged depletions (for a definition see text) for integer (half integer) aspect ratios in the main figure (inset), measuring particle ($D^{(3D)}$, dark blue), spatial transversal ($D^{(s=1,2)}$, light green) and longitudinal ($D^{(s=3)}$, magenta) correlations.
Circles denote MF simulations, and stars represent BMF simulations.
Physical parameters: $N=5$, $H=9$, $S=0.2$, $h=91.125$ and $\sigma=0.1$.
Data points are connected by a line in order to guide the eye.
}
\label{fig_3D_NatPop}
\end{figure}

To quantify particle and spatial correlations, we analyze the integrated depletions  in dependence of
$\eta$, 
\begin{align}
 D^{(3D)}(\eta) &= 1-\frac{1}{T_{max}} \int_0^{T_{max}} a^{(3D)}_1 dt \nonumber \\
 D^{(s)}(\eta)    &= 1-\frac{1}{T_{max}} \int_0^{T_{max}} b_1^{(s)} dt \nonumber
\end{align}
where $a^{(3D)}_1$ and $b_1^{(s)}$ are the first natural populations of $\rho^{(3D)}$ and $\rho^{(s)}$ for the dimension $s\in\{1,2,3\}$\footnote{ Note $D^{(1)}=D^{(2)}$ due to symmetry.}, respectively, and $T_{max}$ is the maximal simulation time.
These quantities may be interpreted as followed:
The larger the averaged depletion $D^{(3D)}(\eta)$, $D^{(s)}(\eta)$ is, the more important correlations are between the atoms and between the spatial dimension $s$ and the other two spatial dimensions in the dynamics, respectively.
Comparing the integer and non-integer valued cases (see figure \ref{fig_3D_NatPop} and its inset), we see that the general characteristics of the depletions are similar, with the only difference that for smaller aspect ratios $\eta <5$, the spatial depletions $D^{(s)}$  are much stronger in the integer valued case.

More precisely, we see that for small aspect ratios the system is spatially correlated but particle correlations are negligible such that the mean-field approximation is applicable.
Increasing the aspect ratios, particle correlations increase while spatial correlations decrease.
In this regime ($\eta \sim 4$), only a BMF simulation in three dimensions, as performed here with the ML-MCTDHB method, can resolve all the correlations playing a substantial role.
Further increasing the aspect ratio, the spatial correlations between the transversal and longitudinal degrees of freedom become negligible and thus the many-body wave function can be approximated first as a product state w.r.t. to the dimensions  and second as a transversally condensed state, where all atoms reside in the same transversal wave function, 
$\Psi(\mathbf{r}_1, \; ... \; \mathbf{r}_N) = [\prod_{i=1}^N \phi(x_i) \phi(y_i) ]\varphi(z_1,\;...\; z_N)$.
This regime can be tackled by a purely one-dimensional approach with the transversal degrees of freedoms being integrated out (see section \ref{section_reduction_1D}).

\begin{figure}[h]
\centering
\includegraphics[width=\linewidth]{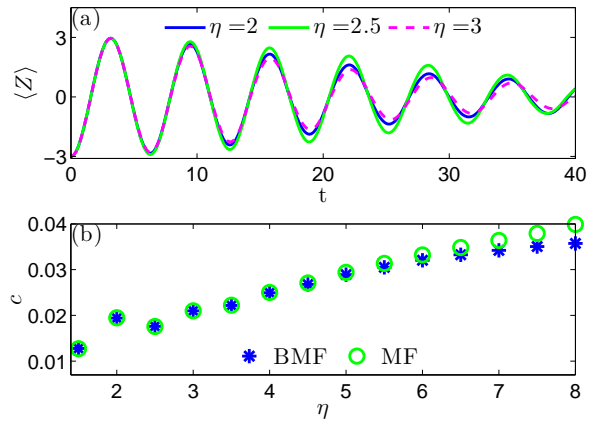}
\caption{ \label{fig_3D_COM}
(Color online)
(a) The CM dynamics is shown for the different aspect ratios $\eta=2, 2.5, 3$. 
(b) shows the fitted decay constant $c$ of the model \eqref{eq_ana_Z} in dependence of $\eta$.
Same physical parameters as in figure \ref{fig_3D_NatPop}.
}
\end{figure}
The CM dynamics of the few-boson ensemble does not differ qualitatively from the single-atom case discussed in section \ref{section_3D_1P} (see figure \ref{fig_3D_COM}a), but only quantitatively due to the presence of interactions.
As already observed, the CM oscillation for integer aspect ratios features a stronger decay, since energy can be stored in transversal modes.
Fitting again the model \eqref{eq_ana_Z} to the numerical data (figure \ref{fig_3D_COM}b), we see that the decay coefficient $c$ features a similar dependence on $\eta$ as for a single atom (figure \ref{fig_1P_3D_COM}), but the peak structure is less pronounced.
Furthermore for larger aspect ratios, a discrepancy in $c$ is observed between MF and BMF simulations.
This effect has already been encountered in the purely one-dimensional case, where the second natural orbital becomes populated, leading to a reduction of the decay coefficient (see section \ref{section_1D_Set-Up}).

\begin{figure}[h]
\centering
\includegraphics[width=\linewidth]{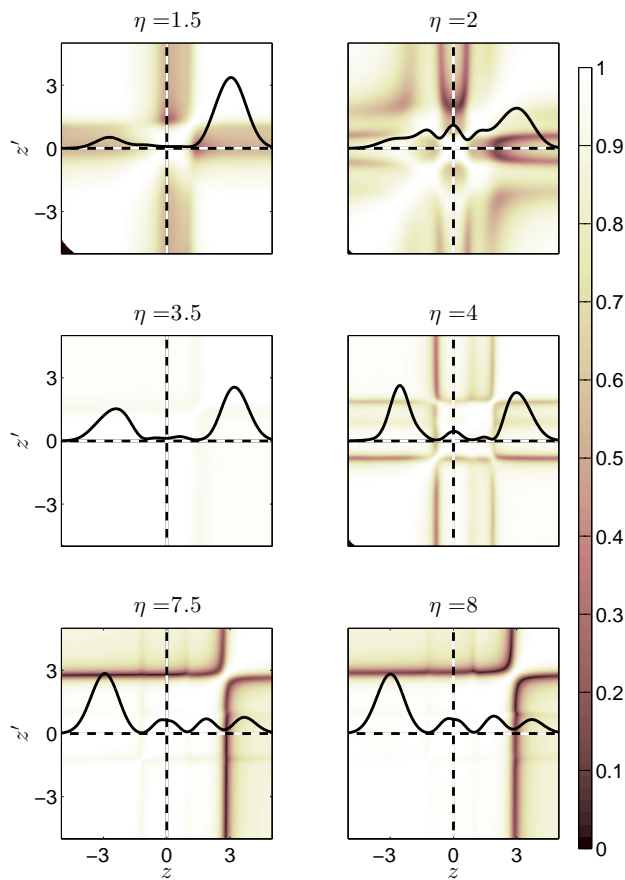}
\caption{ \label{fig_3D_g1}
First-order correlation function $|g_1(z,z')|$ for different aspect ratios $\eta$ at the time instant $t=9\pi$.
Areas of density smaller than $10^{-6}$ are colored in black.
The black-white dashed line marks the position of the barrier and the black solid line is the longitudinal density profile.
Same physical parameters are used as in figure \ref{fig_3D_NatPop}.
}
\end{figure}

In figure \ref{fig_3D_g1}, the absolute value of the first-order correlation functions $g_1(z,z')$ is given for different aspect ratios,  at the right classical turning point ($t=9\pi$).
For the aspect ratios $\eta=1.5, 2$, we see that the correlation function exhibits the characteristic structure observed for a single atom in three dimensions (see figure \ref{fig_1P3D_g1}), namely the loss of coherence between the region close to the barrier and outer regions.
This structure is more pronounced for the integer aspect ratio $\eta=2$ because of the enhanced population of the second transverse excited mode.
Increasing the aspect ratio to $\eta = 3.5, 4$, enhances the overall coherence.
For $\eta=3.5$, we even find almost perfect coherence in the longitudinal direction.
This regime is well-suited for propagating the initial wave function coherently in a harmonic trap with the presence of a scatterer, as a beam splitter.
For large aspect ratios $\eta=7.5, 8$, the differences between the integer and non-integer aspect ratios disappears and an incoherent structure emerges, which is similar to the results for few bosons in one dimensions (see figure \ref{fig_1D_g1}), but with a sharp borderline between coherent regions.

\section{Conclusions} \label{section_conclusion}

In this work, we have analyzed the temporal evolution of an interacting few boson ensemble, initially displaced from the trap center of an elongated harmonic trap.
The bosonic ensemble evolves in time and scatters off a centered barrier.
We explored the change in the amplitude of the oscillation of the center of mass and the loss of first-order coherence in the longitudinal direction due to spatial and particle correlations in dependence on the aspect ratio.
The analysis has been divided into three parts.

First we have been investigating few bosons in one dimension (where only particle correlations are present), second, one boson in three dimensions (where only spatial correlations are present) and finally, five bosons in three dimensions taking particle and spatial correlations into account.

We have identified two mechanisms of coherence loss.
In the first one, present for low aspect ratios, loss of coherence are observed, manifested between positions close to the barrier and outer regions, due to the excitations of transversal modes.
The second one, for larger aspect ratios, is an  incoherence between the density fragments to the right and the left of the barrier, emerging due to particle correlations.
In between, we have found a regime ($\eta=3.5$), where coherent transport of the wave-function is possible, even in the presence of a scatterer and therefore the regime may be suitable to avoid decoherences in beam splitter  and matter-wave interferometers. 

In addition, we have explored the oscillation of the center of mass in dependence of the aspect ratio, which was changed smoothly from three to one dimension. 
Due to multiple scattering off the barrier, each event resulting in a reflected and a transmitted density fraction, the CM oscillation decays.
In addition to this simple mechanism, we have identified two effects influencing the CM dynamics.
First, for weak aspect ratios, the barrier couples the dimensions and energy is transferred into the excitation of transversal modes, leading to a decay mechanism for the CM oscillations, due to energy conservation.
Second, for larger aspect ratios, particle correlations become more pronounced, which reduce the decay of the CM oscillations due to the spatial structure of the second dominant natural orbital and its increasing population.

Furthermore, the above result for small $\eta$ depend on if the aspect ratio is integer or non-integer and has been analyzed.
For integer aspect ratios, the decay of the CM and the loss of coherence are more pronounced in comparison with non-integer aspect ratios, which can be traced back to the constructive or destructive interference of the transversal breathing mode excitations induced by multiple scattering events.

Looking at the depletions,  three different regimes have been identified.
For low aspect ratios, particle correlations are suppressed and the mean-field approximation can be used, whereas for high aspect ratios spatial correlations are reduced and an adiabatic separation can be employed to reduce the three-dimensional to an one-dimensional problem.
In between, both spatial and particle correlations are important and a full ab-initio three-dimensional simulation is needed.

Finally, we note that the unraveled mechanisms may be utilized for state preparation as needed for e.g. interferometric applications.
Tuning the initial displacement, the waiting time, i.e. the number of collisions with the barrier, the aspect ratio to integer or non-integer values allows for controlling the spatial fragmentation into density fragments, their mutual coherence as well as the admixture of excited transversal states.

\section{Acknowledgments}
We thank Johannes Schurer for fruitful discussion.
This work has been supported by the excellence cluster 'The Hamburg Centre for Ultrafast Imaging - Structure, Dynamics and Control of Matter at the Atomic Scale' of the Deutsche Forschungsgemeinschaft.

\appendix

\section{Methodology and computational approach} \label{app_MLMCTDHB}

Before sketching the main concept idea of the ab-initio ML-MCTDHB method, we shortly summarize the numerical challenges, occurring for higher-dimensional interacting bosonic systems (see also \cite{Bolsinger2017} and refs. therein).
At the end of this section, we comment on the convergence behavior of our numerical simulations.

In general, there are two main numerical challenges for non-perturbative methods in three dimensions treating dilute bosonic ensembles.
The first one is the exponential scaling of complexity w.r.t. the number of bosons and the second one is the separation of different length scales (e.g.\ the characteristic interaction and trap length), leading to extremely large grids, and thus an enormous numerical effort, if a product grid is applied.

In order to tackle these numerical challenges, we have recently optimized the ab-initio ML-MCTDHB method \cite{Cao2013, Kronke2013}, for efficiently simulating bosons in three-dimensional traps using huge grids \cite{Bolsinger2017}.
First, we expand the many-body wave function $|\Psi \rangle$ into a set of time-dependent bosonic number states
\begin{equation} 
|\Psi(t)\rangle=\sum_{\vec{n}|N} A_{\vec{n}}(t) |\vec{n}\rangle_{t} \nonumber
\end{equation}
These number states are labeled by an integer vector $\vec{n}=(n_{1},\ ...,\ n_{i},\ ...,\ n_{M})$, where $n_{i}$ is the occupation number of the \textit{$i$}-th three-dimensional, time-dependent single-particle function (3D-SPF), $\left|\chi_{i}(t)\right\rangle $, which is variationally optimized at each instant in time.
The symbol $\left.\vec{n}\right|N$ denotes the summation over all $N$-body number states.
The three-dimensional 3D-SPFs are then expanded w.r.t.\ a product of three one-dimensional, time-dependent single-particle functions (1D-SPFs), $ \{ |\phi_{j_{s}}^{(s)} \rangle \}_{j_s=1}^{m_s} $ with the time-dependent expansion coefficient $B_{ij_1j_2j_3}(t)$
\begin{align}
\left|\chi_{i}(t)\right\rangle & = \sum_{j_1=1}^{m_1} \sum_{j_2=1}^{m_2} \sum_{j_3=1}^{m_3}  B_{ij_1j_2j_3}(t)
|\phi_{j_{1}}^{(1)}(t)\rangle 
|\phi_{j_{2}}^{(2)}(t)\rangle 
|\phi_{j_{3}}^{(3)}(t)\rangle \nonumber
\end{align}
Finally, the 1D-SPFs are expanded w.r.t. some time-independent, one-dimensional basis $ \{ |u^{(s)}_r\rangle \}_{r=1}^{q_s}$, for which we choose a Fast-Fourier transformation based grid \cite{Kosloff1983,Tannor2007} in this work,
\begin{equation}
|\phi_{j}^{(s)}(t)\rangle=\sum_{r=1}^{q_s} C_{jr}^{(s)}(t)|u^{(s)}_r\rangle. \nonumber
\end{equation}
where $s={1,2,3}$ labels the three different dimensions and $C_{jr}^{(s)}(t)$ is the expansion coefficient for the dimension $s$.

The system of equations of motion for the time-dependent coefficients $A_{\vec{n}}(t)$, $B_{ij_1j_2j_3}(t)$ and $C_{jr}^{(s)}(t)$ can be derived \cite{Bolsinger2017}, using e.g. the Dirac-Frenkel variational principle \cite{Dirac1930,Frenkel1932} with the Hamiltonian given in \eqref{eq_3D_Hamilt}.

In doing so, we achieve an additive scaling in the number of grid points $q_s$ w.r.t. the dimensions $s$ for the representation of the wave function, but with the disadvantage of a more involved scaling w.r.t. the number of 1D-SPFs $m_s$ \cite{Bolsinger2017}.
Fortunately, if the correlations between the spatial directions are not too strong, e.g. for elongated traps, where the trap geometry separates the longitudinal and transversal energy scales naturally and where only few transversal modes are populated, the 3D-SPFs can be represented well by taking into account only a few $m_s \ll q_s$ 1D-SPFs.

In the following, we discuss two limiting cases of ML-MCTDHB:
First, if we provide only one 3D-SPF $(M=1)$ and ensure convergence w.r.t. the numbers of 1D-SPFs $m_s$, the ML-MCTDHB equations of motion recover the (mean-field) GPE.
Second, if only one transversal 1D-SPF $(m_1=m_2=1)$ is supplied and convergence w.r.t. $M=m_3$ is ensured, the many-particle wave function adiabatically separates w.r.t. the dimensions, using the same variationally-optimized wave function in the transversal direction for all particles.
In doing so, particle correlations in the longitudinal direction $(s=3)$ can still be resolved.

\paragraph*{Convergence behavior:}
The numerical results depend on seven numerical control parameters; the number of 3D-SPFs $M$, the three numbers of 1D-SPFs $\{m_1, m_2, m_3 \}$ as well as the three numbers of grid points $\{ q_1,q_2,q_3\}$.
We regard a numerical simulation converged, when an observable of interest does not change to a certain desired accuracy, if these numerical control parameters are varied.
In our numerical calculations, we use always a sufficiently large number of grid points, and thus neglect their discussion in the following, reducing the seven dimensional parameter space to a four dimensional one.
In the here performed simulations, we have used $800$ ($200$) grid points for the longitudinal (transversal) direction(s), with a grid spacing of $0.025$.
Due to the symmetry of the elongated trap, we can set $m_1=m_2$.
We call the set of parameters $\mathcal{C}=(M;m_1,m_3)$ a numerical configuration $\mathcal{C}$.

For very strongly elongated traps, where the main dynamics takes place in the longitudinal direction $(s=3)$, we can reduce the three-dimensional parameter space further, by setting  $m_3=M>m_1$, for which case the particle correlations, if existent, are handed over to the population of longitudinal 1D-SPFs.
Whereas, opposite to this case for nearly isotropic trap, the parameter space can be reduced by choosing $m_1=m_2=m_3$.

In the main text, we have mainly focused on two observables, first, the oscillation of the CM $\langle Z \rangle$ and, second, the occupation of the first natural orbital $a_1$.
Their convergence is exemplary shown now by regarding the integrated difference between two numerical configurations, $\mathcal{E}_1 = \int_0^{T_{max}} |\langle Z \rangle _{\mathcal{C}_1}-\langle Z \rangle _{\mathcal{C}_2}| dt / T_{max}$ and $\mathcal{E}_2 = \int_0^{T_{max}} |a_{1,\mathcal{C}_1}-a_{1,\mathcal{C}_2}| dt / T_{max}$, where the subindex denotes the used numerical configuration $\mathcal{C}$ and $T_{max}$ is the maximal simulation time.
We compare the numerical configuration $\mathcal{C}_1=(5;3,5)$ and $\mathcal{C}_2=(6;4,6)$ for the same physical parameters as used in section \ref{section_3D} both for $\eta=2$ and $\eta=8$, which are the extreme cases for spatial and particle correlation respectively.
For $\eta=2$, we obtain $\mathcal{E}_1=5.8\cdot 10^{-3}$ and $\mathcal{E}_2=6.1 \cdot 10^{-4}$ as well as for $\eta=8$ we get $\mathcal{E}_1=3.1\cdot 10^{-3}$ and $\mathcal{E}_2=2.2 \cdot 10^{-3}$.
In essence, the integrated error is estimated to be of the order $10^{-3}$.

\section{Co-moving time-dependent basis states} \label{section_1D_TDBF}

In the first part of this appendix, we derive a complete set of orthonormal functions $\phi_n(z,t)$, described by displaced harmonic oscillator functions, stiffly\footnote{see footnote $\ref{footnote1}$} oscillating in a harmonic trap, which are used in section \ref{section_3D_1P}.
In the second part of this section, we show that even the displaced stationary ground state of  $\varphi_{GP}(z)$ (see equation \eqref{eq_1D_TDGPE}) performs stiff oscillations in a harmonic trap as well \cite{Moulieras2012}, as utilized in section \ref{section_1D_results}.

First, the orthonormal functions $\phi_n(z,t)$ are assumed to be of the following functional form with the yet unknown real-valued functions $\Theta_n(t)$, $\bar{p}(t)$ and $\bar{z}(t)$.  
\begin{equation} \label{eq_1D_set_of_functions}
  \phi_n(z,t) = e^{-i \Theta_n(t) +i \bar{p}(t)z } \varphi^{1D}_n(z-\bar{z}(t))  
\end{equation}
where $\varphi^{1D}_n$ is the  $n$-th harmonic oscillator function $\varphi^{1D}_n(x) = 1/\sqrt{2^n n !} \pi^{-1/4} \exp \left( -x^2/2 \right) H_n \left( x\right)$ with the Hermite polynomials $H_n$.
The ansatz \eqref{eq_1D_set_of_functions} is inserted into the time-dependent, one-dimensional, single-particle Schr\"odinger equation, $i\partial_t \phi_n = H^{(1)}_{0,z} \phi_n$,  and we obtain three coupled differential equations by comparing the real and imaginary part as well as equating coefficients
\begin{align} \label{eq_set_of_coupled_eq}
   \partial_t \bar{z}(t) &= \bar{p}(t) \\ \nonumber
 - \partial_t \bar{p}(t) &= \bar{z}(t) \\ \nonumber
   \partial_t \Theta_n(t) - \bar{z}(t) \partial_t \bar{p}(t)   &= E_n + \frac{1}{2} \left( \bar{z}^2(t) + \bar{p}^2(t)\right)   \nonumber
\end{align}
with $E_n=n+1/2$.
With the initial condition that the wave functions is displaced by $b$,  $\phi_n(z,0)=\varphi^{1D}_n(z-b)$, the coupled set of equations can be solved 
\begin{align} \label{eq_solutions_of_coupled_eq}
   \bar{z}(t) &= b \cos(t) \\ \nonumber
   \bar{p}(t) &= - b \sin(t) \\
   \Theta_n(t)  &= E_n t + \frac{1}{2} \bar{z}(t) \bar{p}(t)   \nonumber
\end{align}
The functions $\phi_n(z,t)$ form a complete and orthonormal set of basis functions at all instants in time.
Orthonormality can be checked easily and the proof of completeness follows the same arguments as for the Hermite polynomials \cite{Hochstadt1971}.

Second, the initially displaced mean-field ground-state wave functions $\varphi_{GP}$ performs also stiff oscillations in and only in a harmonic trap \cite{Moulieras2012}.
The ground state mean-field orbital obeys $E_{GP} \varphi_{GP}(z) = \left( -1/2 \partial_z^2 + 1/2 z^2 \right) \varphi_{GP}(z) + g(N-1) \int d\mathcal{Z} |\varphi_{GP}(\mathcal{Z})|^2  W(z,\mathcal{Z}) \varphi_{GP}(z)$.
Inserting the same approach for the wave function $\phi_{GP}(z,t) = e^{-i \Theta(t) +i \bar{p}(t)z } \varphi_{GP}(z-\bar{z}(t))$ into the corresponding time-dependent GPE, where $W(z_1,z_2)=W(z_2-z_1)$ is assumed, leads again to the three coupled differential equations \eqref{eq_set_of_coupled_eq} with their solution \eqref{eq_solutions_of_coupled_eq}, but now with the energy $E_n$ replaced by $E_{GP}$.

%

\end{document}